\documentclass[onecolumn,times]{aastex62}

\usepackage{amsmath}
\usepackage{lineno}

\received{\today}
\revised{\today}
\accepted{\today}
\shortauthors{Hayes et al.}
\usepackage[printonlyused]{acronym}

\begin{document}

\title{Comparing Short Gamma-Ray Burst Jet Structure Models}

\correspondingauthor{Fergus Hayes}
\email{f.hayes.1@research.gla.ac.uk}

\author{Fergus Hayes}
%\author{\fergus{Fergus Hayes}}
\affiliation{SUPA, School of Physics and Astronomy,
University of Glasgow\\
Glasgow G12 8QQ, United Kingdom}

\author{Ik Siong Heng}
%\author{\siong{Ik Siong Heng}}
\affiliation{SUPA, School of Physics and Astronomy,
	University of Glasgow\\
	Glasgow G12 8QQ, United Kingdom}

\author{John Veitch}
%\author{\jv{John Veitch}}
\affiliation{SUPA, School of Physics and Astronomy,
	University of Glasgow\\
	Glasgow G12 8QQ, United Kingdom}

\author{Daniel Williams}
%\author{\dw{Daniel Williams}}
\affiliation{SUPA, School of Physics and Astronomy,
	University of Glasgow\\
	Glasgow G12 8QQ, United Kingdom}



\begin{abstract}
A structured gamma-ray burst jet could explain the dimness of the prompt emission observed from GRB$\,170817$A but the exact form of this structure is still ambiguous. 
However, with the promise of future joint gravitational wave and gamma-ray burst observations, we shall be able to examine populations of binary neutron star mergers rather than a case-by-case basis. 
We present an analysis that considers gravitational wave triggered binary neutron star events both with and without short gamma-ray burst counterparts assuming that events without a counterpart were observed off-axis. 
This allows for Bayes factors to be calculated to compare different jet structure models. 
We perform model comparison between a Gaussian and power-law apparent jet structure on simulated data to demonstrate that the correct model can be distinguished with a log Bayes factor of $>5$ after less than 100 events.
Constraints on the apparent structure jet model parameters are also made. After 25(100) events the angular width of the core of a power-law jet structure can be constrained within a $90\%$ credible interval of width $ \sim9.1(4.4)^{\circ} $, and the outer beaming angle to be within $\sim19.9(8.5)^{\circ}$.
Similarly we show the width of a Gaussian jet structure to be constrained to $\sim2.8(1.6)^{\circ}$.

%\fergus{250 words limit. Needs adding to once new results come through}

\end{abstract}



\section{Introduction} \label{sec:intro}

%\dw{You might want to consider using the acro package for handling the large number of acronyms in the introduction.}

Gamma-ray bursts (GRBs) are pulses of high energy electromagnetic (EM) radiation from astrophysical sources. 
There is strong evidence that there are two populations of GRBs --- long GRBs and short GRBs (sGRBs)~\citep{kouveliotou1993identification}. 
sGRBs are generally shorter in duration and have harder spectral energies. 
It has been long believed that the merger of binary neutron star (BNS) systems are the progenitors of sGRBs, and in August 2017 this was confirmed by the detection of GRB$\,170817$A~\citep{goldstein2017ordinary,savchenko2017integral}. 
The observation was accompanied by a gravitational wave (GW) compact binary coalescence signal, GW170817, by the LIGO and VIRGO scientific collaboration~\citep{abbott2017gw170817} and therefore became the first joint EM and GW observation~\citep{abbott2017gravitational} --- a landmark in multimessenger astronomy. 
However, while GRB$\,170817$A was the closest detected sGRB, at a distance of $\sim 40$~Mpc, it was also $ \sim10^{3} $ times less energetic than any of the weakest previously observed sGRBs with known redshift.
While there are numerous proposed explanations, it is thought that either sGRB 
energies fall below the current assumed energy range, that GRB$\,170817$A was 
viewed off-axis or that sGRBs have a non-uniform energy distribution within the solid angle subtended by their jets.

There had been much work on the jet structure of long GRBs prior to the detection of GRB$\,170817$A; \citet{rossi2002afterglow}, \citet{zhang2002gamma} and further work by \citet{rossi2004polarization}, \citet{zhang2004quasi} and \citet{lloyd2004structure} consider a \textit{quasi-universal} long GRB jet structure, where discrepancies in observed isotropic energies across observations is due to the varying viewing angle of events described by the same underlying jet structure and show that a Gaussian or power-law jet structure can each fully account for all variations in observed long GRBs until that date. 
\citet{lazzati2005universal} consider the effect of surrounding shocked matter about the jet, known as a jet cocoon, as a possible cause for a quasi-universal jet structure. 
Relativistic magnetohydrodynamic simulations have shown a structured jet can naturally form without a cocoon~\citep{aloy2005relativistic,kathirgamaraju2019counterparts}.
It has also been shown that radial and angular structuring can evolve from jets that are initially uniform in energy distribution through hydrodynamical interactions with surrounding material in numerical simulations~\citep{xie2018numerical,gill2019numerical}.

Determining the true jet-energy distribution will allow both an explanation for
the luminosity of GRB$\,170817$A, as well as insight into the astrophysics behind the 
jet formation caused by the BNS merger.
This energy distribution may be described by a shape function $y(\theta)$, 
where $\theta$ is the polar angle from the jet axis such that $0\le\theta\le\pi/2$.
Although the electromagnetic flux at the detector is known (which is proportional to $y$), in 
order to infer the distribution we need some way of measuring the viewing angle $\theta_{\text{v}}$. 
For simplicity, we may assume that a universal jet structure exists 
so that by observing sGRBs with various $\theta_{\text{v}}$ and $y$ values we can 
build up a picture of the jet structure.
$\theta_{\text{v}}$ may be measured by radio observations of the superluminal motion of the jet over the months following the event \citep{mooley2018superluminal}, by measuring the source size over time~\citep{ghirlanda2019compact}, as well as from the spectral features of kilonovae \citep{kasen2015kilonova,metzger2017kilonovae}.
GW signals can be used to infer the joint probability 
distribution of source parameters, including both the luminosity distance $d_{\text{L}}$ 
and inclination angle $\cos\iota=\vec{J}\cdot\vec{N}$ where $ \vec{J} $ is the angular momentum vector of the system and $ \vec{N} $ is between the system and the observer.
Under the assumption that all sGRBs are generated by BNS mergers, this will give us 
the information needed to measure the energy emission (from the EM flux and 
luminosity distance) and the viewing angle $\theta_{\text{v}} = \mathrm{min}(\iota, 
\pi-\iota)$.
By observing multiple BNSs with both EM and GW channels, and 
assuming a common jet model, we can gradually build up a picture of the apparent jet 
energy distribution function.
This is even the case for BNSs that are only detectable from their GWs and without detectable sGRBs; if we can deduce that the sGRB was undetectable due to being viewed outside the confines of the jet the inferred viewing angle can be used to constrain the width of the jet structure.
We derive a method to implement this idea, demonstrate it in a simulated 
dataset and investigate the number of events required to measure jet parameters 
and perform model selection.
Using future joint GW and prompt sGRB detections to investigate the underlying jet structure has been explored in previous work.

\citet{beniamini2018lesson} and \citet{gupte2018observational} consider different jet structures to infer the possible rates of joint GW and prompt GRB detections, an approach that has recently been expanded upon by \citet{farah2019counting}.
These studies, as well as this work, consider the \textit{apparent} jet structure of sGRBs rather than the \textit{intrinsic} jet structure (see Section~\ref{sec:jetstructs} for details on the difference between intrinsic and apparent jet structures).
Recent work by \citet{biscoveanu2019constraining} also considers a similar analysis to this work, but uses the joint GW and sGRB detections to probe the intrinsic jet structure, however we present the first use of this method to compare different jet structure models.

In Section~\ref{sec:jetstructs}, we specify the Gaussian and power-law apparent jet structure models that are compared in this work. 
In Section~\ref{sec:model}, we discuss the model and state the likelihood and 
priors used in analyzing the EM and GW data.
Section~\ref{sec:results} contains the results when given sets of 100 simulated BNS events of either jet structure model.
Lastly, we further discuss the results and conclude in Section~\ref{sec:conclusion}.

\section{Jet Structures}\label{sec:jetstructs}

Jet structuring describes the sGRB jet energy distribution over solid angle.
We assume that the distribution is axisymmetric and therefore can be described only in terms of a function of the polar angle, $\theta$.
\citet{salafia2015structure} introduce the concept of \textit{apparent} and \textit{intrinsic} jet structure which is adopted here.
The intrinsic structure is the angular energy distribution in the frame of the event while the apparent structure is dependent on the observer's frame and so are related by a Lorentz transformation.
The difference between these two distributions is explained by 
\textit{relativistic beaming}: the apparent area of the source that the 
observer receives radiation from depends on the Lorentz factor
$\Gamma$ of the jet and the viewing angle, $\theta_{\text{v}}$.
When $\Gamma\gtrsim1$, the observer receives emission from the whole visible 
emitting surface of the source (the head of the jet).
This can lead to emission being observed even when $\theta_{\text{v}}$ is outside the 
confines of the intrinsic jet structure - often termed an off-axis 
observation~\citep{granot2002off}. 
In the ultrarelativistic limit, with $\Gamma\gg1$, the observed prompt radiation is mostly from the emitting jet material traveling along the observers line of sight $\theta=\theta_{\text{v}}$, and little from any surrounding jet material~\citep{rhoads1997tell}.
Therefore the intrinsic jet structure depends on both the angular Lorentz factor and energy distribution in the frame of the event.
Additionally \citet{beniamini2018observational} demonstrate that even with a narrow Lorentz factor distribution low gamma-ray production efficiency at large angles suppresses emission and causes narrow jets for long GRBs.

For sources at negligible redshifts, the apparent isotropic equivalent energy of an event viewed at angle $\theta_{\text{v}}$ can be written as:

\begin{eqnarray}\label{equ:Eiso}
E_{\text{\text{iso}}}(\theta_{\text{v}})=4\pi d_{\text{L}}^{2} \mathcal{F},
\end{eqnarray}

where $\mathcal{F}$ is the observed fluence. 
For BNS mergers close enough for GW detection ($\lesssim400~$Mpc), cosmological effects are small. 
Therefore we assume there is negligible redshifting and that the source spectrum is fully apparent to the detectors, removing the necessity for a cosmological $k$-correction \citep{bloom2001prompt}.
We describe $E_{\text{\text{iso}}}(\theta_{\text{v}})$ in terms of the face-on isotropic equivalent 
energy $E_{\text{iso},0}=E_{\text{\text{iso}}}(0)$ and shape function of the apparent 
structure $y(\theta)$:
\begin{eqnarray}\label{equ:shapeE}
E_{\text{\text{iso}}}(\theta_{\text{v}}) = E_{\text{iso},0}\,y(\theta_{\text{v}})\quad\text{ where 
}\quad y(0)=1. 
\end{eqnarray}

\begin{figure*}
	\gridline{\fig{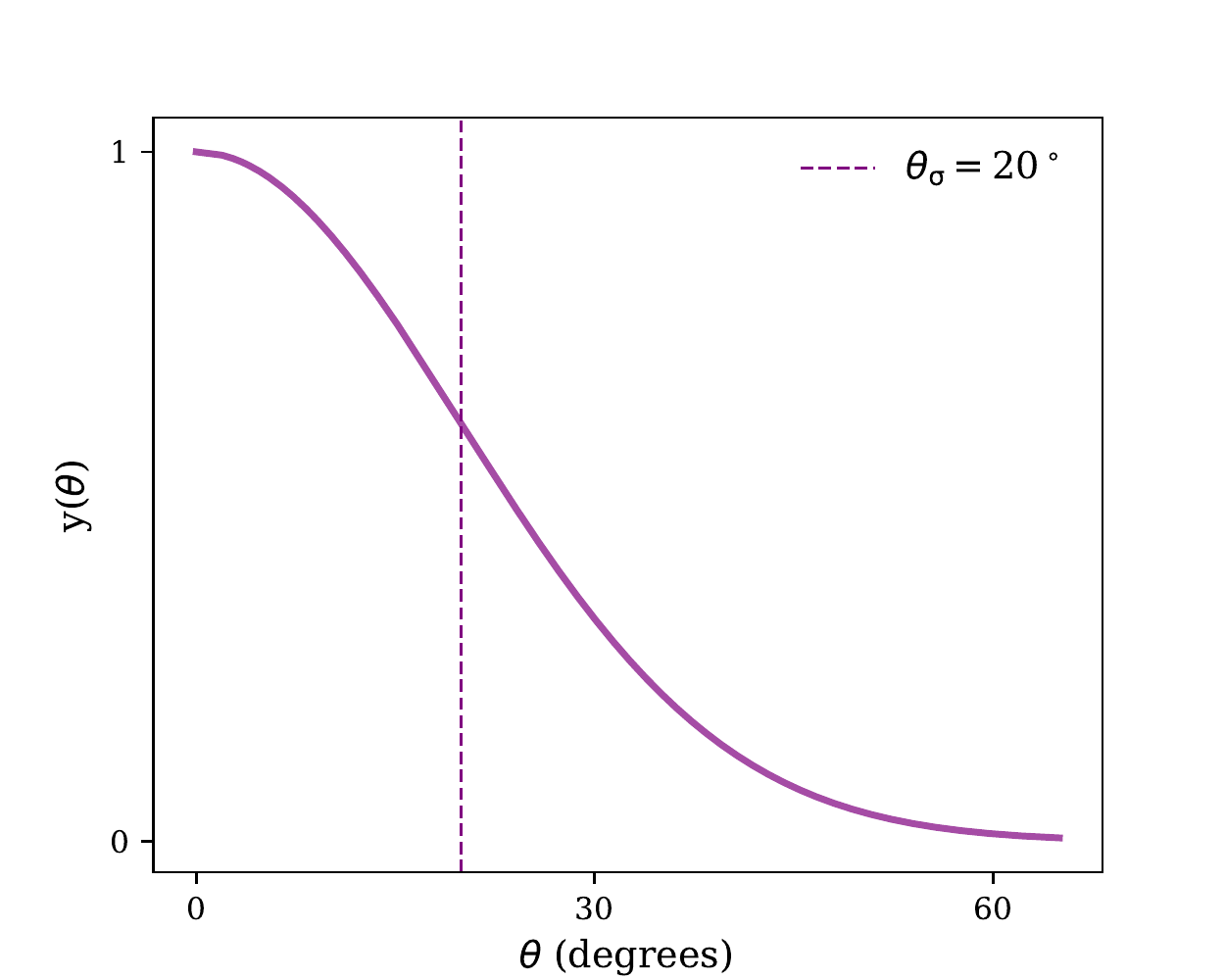}{0.5\textwidth}{(a) Gaussian jet \label{fig:y_gauss}}
		\fig{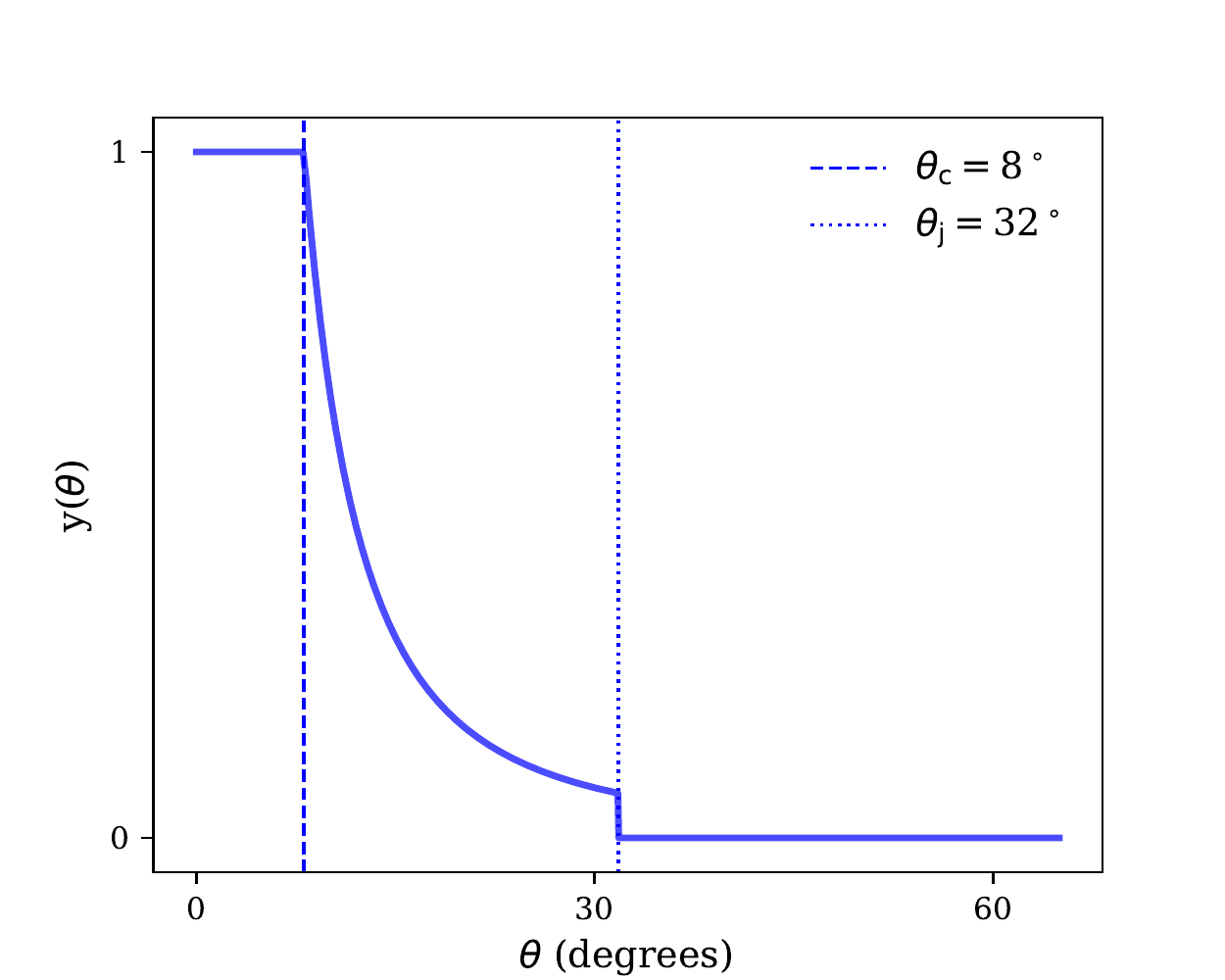}{0.5\textwidth}{(b) Power-law jet \label{fig:y_plaw}}}
	\caption{Functional forms of jet structure models. (Left) The Gaussian jet structure parameterized by width $\theta_{\sigma}=20^{\circ}$ (dashed) as defined in Equation~\ref{equ:GB}. (Right) The power-law jet structure parameterized by $\theta_{\text{c}}=8^{\circ}$ (dashed) and $\theta_{\text{j}}=32^{\circ}$ (dotted) as in Equation~\ref{equ:PL}. \label{fig:y}}
\end{figure*}

In this proof-of-principle analysis we concentrate on two simple inhomogeneous models to describe $y(\theta)$: a 
Gaussian jet, and a power-law jet, whose apparent structure functions are shown 
in 
Figure~\ref{fig:y}.
These models are often used to describe the intrinsic jet structure of long and short GRBs \citep{zhang2002gamma,rossi2004polarization,kumar2003evolution,lloyd2004structure,salafia2015structure,lamb2017electromagnetic,beniamini2018lesson,oganesyan2019structured,mogushi2019jet}.
Assuming the jet is ultrarelativistic with a bulk Lorentz factor $\Gamma>100$
with little variation across $\theta$ and constant gamma-ray production efficiency, emission from off-axis observations 
is less significant, and the apparent structure closely resembles the 
intrinsic structure \citep{salafia2015structure}. 
Therefore we adopt these models and use them to approximate the apparent jet 
structure in this analysis, which is appropriate for the dominant on-axis 
emission.

\paragraph{Power law jet} The power-law jet describes a structure with a 
uniform energy distribution until angle $\theta_{\text{c}}$, after which it decays as a 
power law $ (\theta/\theta_{\text{c}})^{-k} $ with gradient $k$. 
This model was initially proposed by \citet{meszaros1998viewing} to explain the power-law fit to the decay in the afterglow light curve. 
Initial work in \citet{rossi2002afterglow,zhang2002gamma,lazzati2005universal} suggested a value of $k \sim 2$ for long GRBs to hold with the observed relation 
$E_{\text{\text{iso}}}\propto\theta_{\text{v}}^{2}y(\theta)=~constant$ from \citet{frail2001beaming} but 
\citet{pescalli2015luminosity} shows that $k>2$ allows for a better fit to 
data. 
However, like \citet{lamb2017electromagnetic} and \citet{oganesyan2019structured} we assume $k=2$ for simplicity. 
The power-law jet can be further parameterized by including a sharp cut-off at opening angle $\theta_{\text{j}}$ where $\theta_{\text{c}}<\theta_{\text{j}}$.
This is supported by simulations \citep{aloy2005relativistic,rezzolla2011missing}.

The power-law shape function is then:

\begin{eqnarray}\label{equ:PL}
y_{\text{PL}}(\theta) =
\begin{cases} 
1 & \text{if } \ 0 \le \theta \le \theta_{\text{c}}, \\
\left(\frac{\theta}{\theta_{\text{c}}}\right)^{-2} & \text{if } \ \theta_{\text{c}} < \theta \le \theta_{\text{j}}, \\
0 & \text{if } \ \theta_{\text{j}} < \theta.
\end{cases}
\end{eqnarray}

\paragraph{Gaussian jet} The Gaussian jet depends on a width parameter 
$\theta_\sigma$ such that:

\begin{eqnarray}\label{equ:GB}
y_{\text{GJ}}(\theta) = e^{-\frac{1}{2}\left(\frac{\theta}{\theta_{\sigma}}\right)^{2}}.
\end{eqnarray}

The Gaussian jet structure was initially proposed in \citet{zhang2002gamma} as an alternative quasi-universal jet structure model that could explain the relation found in \citet{frail2001beaming}. Since then, Gaussian-like jet structures have been reproduced in simulations \citep{mckinney2006general}. \citet{lyman2018optical,troja2018outflow,lamb2018grb} later found that a Gaussian jet is preferable to a power-law jet structure in fitting the broad-band GRB$\,170817$A afterglow data. 

\section{The Model} \label{sec:model}

We consider a dataset $\boldsymbol{\mathcal{D}}$ consisting of the GW strain $\boldsymbol{\mathcal{H}}$ and corresponding measured fluence $\boldsymbol{\mathcal{F}}$ measurements of $N$ BNS merger 
events, where $\boldsymbol{\mathcal{D}}=\{\boldsymbol{\mathcal{F}},\boldsymbol{\mathcal{H}}\}$.
The analysis is described by parameters $\boldsymbol{\Theta}$ and assumes a jet 
structure model $M$ while $I$ denotes information used to set the priors and 
form of the likelihood functions.
The parameters describing the EM emission in the jet are $ 
\boldsymbol{\Theta}=\{ \boldsymbol{\theta}_{M},\boldsymbol{E}_{\text{iso},0} \} $, which is the 
set of structure parameters of model $M$ and the $E_{\text{iso},0}$ for each event 
respectively. 
While ignoring the evidence, the posterior of $\boldsymbol{\Theta}$ can be written:

\begin{eqnarray*}
p(\boldsymbol{\Theta}|\boldsymbol{\mathcal{D}},M,I) \propto \int p(\boldsymbol{\mathcal{D}}|\boldsymbol{\Theta},\boldsymbol{\Phi},M,I)p(\boldsymbol{\Theta},\boldsymbol{\Phi}|M,I)d\boldsymbol{\Phi},
\end{eqnarray*}

where $\boldsymbol{\Phi}$ are the set of nuisance parameters, consisting of the luminosity distance and viewing angle of each event, $\boldsymbol{\Phi}=\{\boldsymbol{d}_{\text{L}},\boldsymbol{\theta_{\text{v}}}\}$ that are marginalized over. 
This integral can be approximated from $S$ samples of the joint posterior distributions on $\boldsymbol{d}_{\text{L}}$ and $\boldsymbol{\theta_{\text{v}}}$ produced from GW inference:

\begin{eqnarray}\label{equ:moneyshot}
p(\boldsymbol{\Theta}|\boldsymbol{\mathcal{D}},M,I) \propto \frac{1}{S} \sum_{i=1}^{S} p(\boldsymbol{\mathcal{F}}|\boldsymbol{\Theta},\boldsymbol{\Phi}^{i},M,I)p(\boldsymbol{\Theta}|M,I) \textrm{ where } \boldsymbol{\Phi}^{i} \sim p(\boldsymbol{\Phi}|\boldsymbol{\mathcal{H}},I).
\end{eqnarray}

	\subsection{GRB Likelihood}

For each event, it is assumed that the measured fluence $\mathcal{F}$ is equal to the fluence from the sGRB $\mathcal{F}_{\mu}$ with some added background fluence $b$ and Gaussian noise, which is assumed to be drawn from a normal distribution with standard deviation $\sigma_{\text{n}}$, which is not constant over events. 
Therefore the GRB likelihood is the product of all events:

\begin{eqnarray}\label{equ:lnL}
p(\boldsymbol{\mathcal{F}}|\boldsymbol{\Theta},\boldsymbol{\Phi},M,I) = \prod_{j=1}^{N}\frac{1}{\sqrt{2\pi\sigma_{\text{n}}^{j}}} \exp\left[-\frac{1}{2}\left(\frac{\mathcal{F}^{j}-(\mathcal{F}_{\mu}(\Phi^{j},\Theta^{j},M,I)+b^{j})}{\sigma_{\text{n}}^{j}}\right)^{2}\right].
\end{eqnarray}

Values $b$ and $\sigma_{\text{n}}$ are assumed to be known. 
$\mathcal{F}_{\mu}$ depends on each event parameters and the assumed model and can be calculated using Equations~\ref{equ:Eiso} and \ref{equ:shapeE}.

	\subsection{Priors}\label{sec:priors}

Priors are placed on $E_{\text{iso},0}$ for each event and the model dependent structure parameters $\theta_{M}$. 
The prior on $E_{\text{iso},0}$ is independent of the assumed jet model and is a log-normal distribution \citep{frail2001beaming,salafia2015structure}: 

\begin{eqnarray}\label{equ:Egen}
p(\ln (E_{\text{iso},0}/4\pi)) = \mathcal{N}(\mu_{\tiny \text{E}},\sigma_{\tiny \text{E}}),
\end{eqnarray}

with a mean $\mu_{\tiny \text{E}}=\ln10^{49}$ and standard deviation $\sigma_{\tiny \text{E}}=1$.
This is a rather narrow energy prior, but we assume most of the variation in the GRB isotropic equivalent energy is due to the jet structuring.
For the Gaussian jet, $p(\theta_\sigma|M,I)=(\pi/2)^{-1}$ with 
$0<\theta_\sigma<\pi/2$. For the power-law jet, we use a uniform prior 
$p(\theta_{\text{j}}|M,I)=(\pi/2)^{-1}$ and $p(\theta_{\text{c}}|\theta_{\text{j}},M,I)=\theta_{\text{j}}^{-1}$, 
with $\theta_{\text{c}} < \theta_{\text{j}}$.

\section{Data Simulations} \label{sec:data}

The model is tested on simulated BNS mergers data. 
A value for $ d_{\text{L}} $, $ \theta_{\text{v}} $ and $ E_{\text{iso},0} $ is assigned to each event. 
Events are assumed to be uniformly distributed within the space constrained by the detector horizon, $ d_{\text{max}} = 400\,$Mpc, such that $p(d_{\text{L}}) = 3d_{\text{L}}^{2}/d_{\text{max}}^{3}$. The $\theta_{\text{v}}$ of each event is assumed to be distributed isotropically, and therefore $p(\cos\theta_{\text{v}}) = \mathcal{U}\left(0,1\right)$. Each 
$E_{\text{iso},0}$ is sampled from the prior distribution described in Equation~\ref{equ:Egen}. 
The observed GRB fluence of each event is then generated from these variables from the likelihood in Equations~\ref{equ:lnL} by assuming a jet structure model and assigning values to each $\theta_M$. 
The injected model parameters are $\theta_{\sigma}=20^{\circ}$, $\theta_{\text{c}}=8^{\circ}$ and $\theta_{\text{j}}=32^{\circ}$.
The posteriors on $d_{\text{L}}$ and $\cos\theta_{\text{v}}$ are approximated by bivariate normal distributions with a covariance found from averaging fitted normal distributions to GW posteriors analyzed with \textsc{LALInference} --- a software library for performing Bayesian inference on compact binary coalescence signals~\citep{veitch2015parameter}. 
The GW posteriors are produced from simulated BNS injections given to the advanced LIGO and advanced Virgo network at 2019 sensitivity with a network signal-to-noise ratio $>8$. 

The fluence parameters $\sigma_{\text{n}}$ and $b$ would vary over detections as well between different detectors when considering real events, where they would then have to be determined on a case-by-case basis.
For example, given the merger time of a BNS merger signal from a GW detection, we can search for a corresponding sGRB in data measured by \textit{Fermi's} Gamma-ray Burst Monitor (GBM) using a technique similar to that in \citet{blackburn2015high} and \citet{burns2019fermi} where the background can be fitted using the method described in \citet{goldstein2016updates} and the fluence can be determined by fitting spectral models as described in \citet{gruber2014fermi} and \citet{yu2016fermi}.
For these simulations, the standard deviation on $\mathcal{F}$ scales with the signal strength and background so that $\sigma_{\text{n}}=\sqrt{k(\mathcal{F}_{\mu}+b)}$, to approximate Poisson statistics where we have assumed that the photon count scales with the fluence linearly with constant $k$ which is dependent on the detector's effective area and energy band.
For these simulations we assume all photons in the 10-300~keV energy band have energies $\sim150\,$keV and are received by a detector with surface area of $\sim 300\,\text{cm}^{2}$ so that $k=7\times10^{-10}\,\text{erg}\,\text{cm}^{-2}\,\text{photon}^{-1}$. 
The background is set to be constant across events to $b=3.7\times10^{-6}\,\text{erg}\,\text{cm}^{-2}$, which is an approximation of the background for GRB$\,170817$A integrated over the burst duration for energies ranging from $ 10-300\, $keV~\citep{goldstein2017ordinary}.
In reality, the background count rate of the GBM fluctuates over short timescales and a large background count rate increases uncertainties on the fluence measurements of weak sGRB signals.

\section{Results} \label{sec:results}

We test the analysis on two datasets, each of $N=100$ BNS events that are simulated as described in Section~\ref{sec:data}, each with $S=50$ samples from the joint $d_{\text{L}}$ and $\theta_{\text{v}}$ posteriors. 
One dataset $\mathcal{D}_{\text{GJ}}$ consists of sGRBs produced from the Gaussian jet structure model and the other dataset $\mathcal{D}_{\text{PL}}$ that from a power-law jet structure model. 
This is then repeated for 4 realizations of $\mathcal{D}_{\text{GJ}}$ and $\mathcal{D}_{\text{PL}}$.

PyMC3's NUTS sampler \citep{hoffman2014no} is used to calculate the posteriors in Equation~\ref{equ:moneyshot} and the marginal likelihoods in Equation~\ref{equ:modelcompare}.  

\begin{figure}[h!]
	\plotone{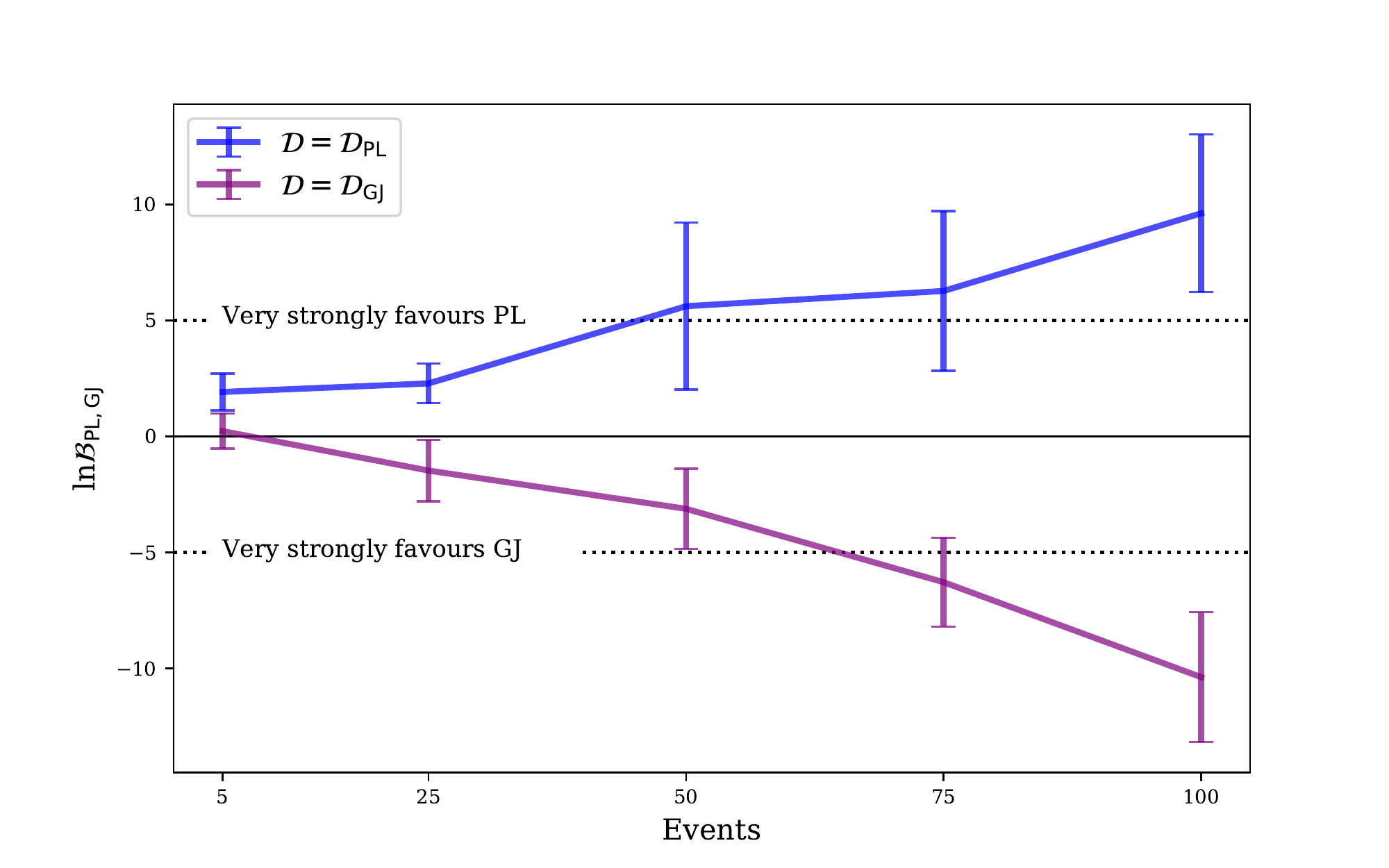}
	\caption{The mean $\ln \mathcal{B}_{\text{PL,GJ}}$ with $1\sigma$ error bars evaluated on supersets of increasing number of events from 4 instances of $\mathcal{D}_{\text{PL}}$ (blue) and $\mathcal{D}_{\text{GJ}}$ (purple) datasets. 
	Values of $\ln \mathcal{B}_{\text{PL,GJ}}$ increase with more events, given the $\mathcal{D}_{\text{PL}}$ while conversely decreasing when given more events of $\mathcal{D}_{\text{GJ}}$, thus allowing for the models to be distinguished with increasing confidence.\label{fig:model_comp_4}}
\end{figure}

For each dataset $\ln\mathcal{B}_{\text{PL,GJ}}$ is calculated according to Appendix~\ref{sec:TI}, as well as for subsets of these datasets. These are shown in Figure~\ref{fig:model_comp_4} where the two lines plot the mean $\ln\mathcal{B}_{\text{PL,GJ}}$ over the 4 different realizations for the subsets of $\mathcal{D}_{\text{PL}}$ in blue and $\mathcal{D}_{\text{GJ}}$ in purple. 
The error bars show $1\sigma$ standard deviation between the realizations
$\ln\mathcal{B}_{\text{PL,GJ}}$ is defined by Equation~\ref{equ:lnBF} so that a positive value indicates the data is best described by the power-law jet structuring model, while a negative value implies the Gaussian jet structuring model. 
The magnitude of $\ln\mathcal{B}_{\text{PL,GJ}}$ is proportional to the certainty the analysis has of this decision. 
Figure~\ref{fig:model_comp_4} demonstrates the analysis can successfully identify the correct underlying model with increasing confidence as more events are considered.

\begin{figure}[h!]
	\gridline{\fig{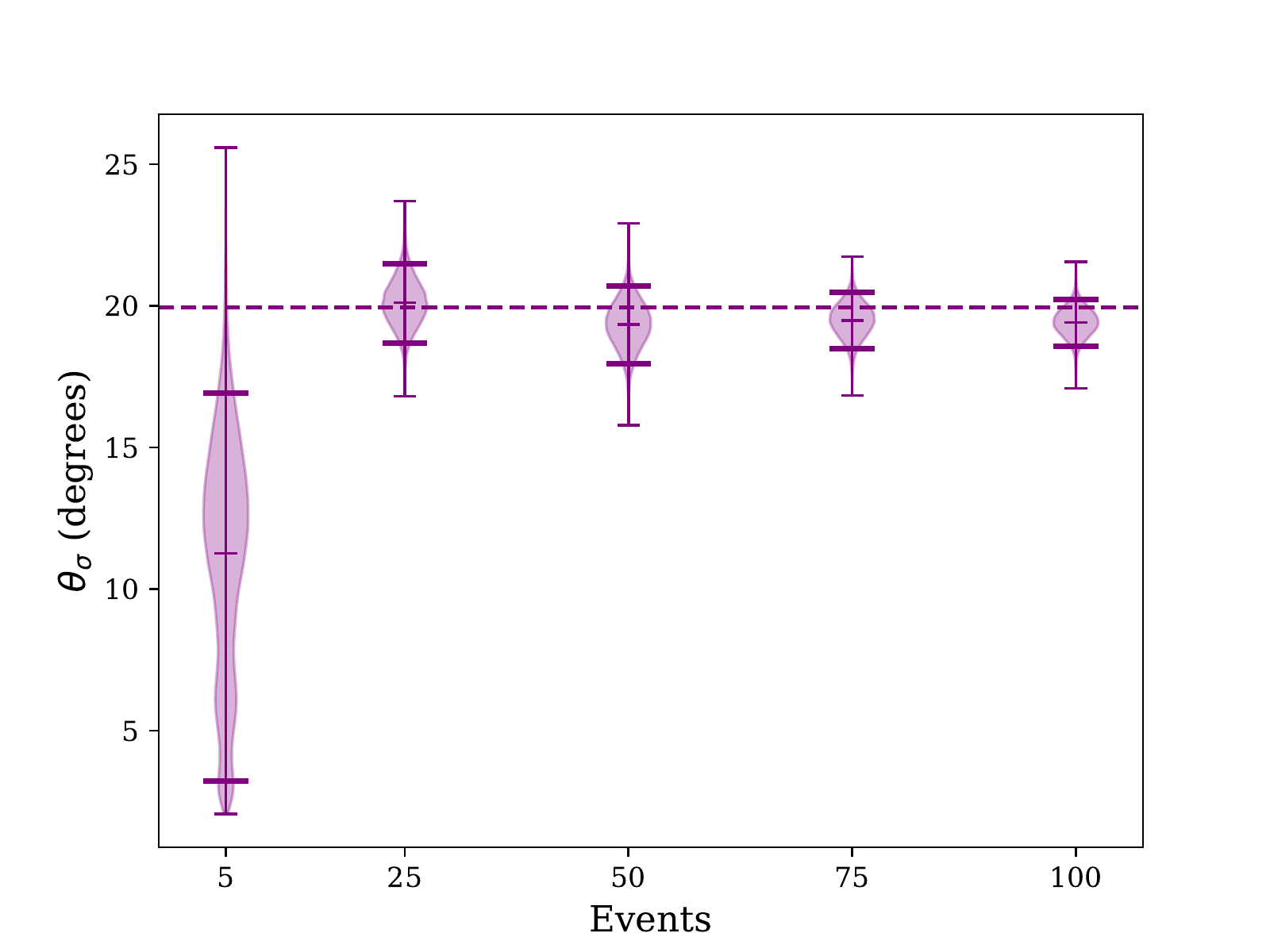}{0.5\textwidth}{(a) $p(\theta_{\sigma}|\mathcal{D}_{\text{GJ}}^{\ast},M=\text{GJ},I)$ \label{fig:GB_violins}}
		\fig{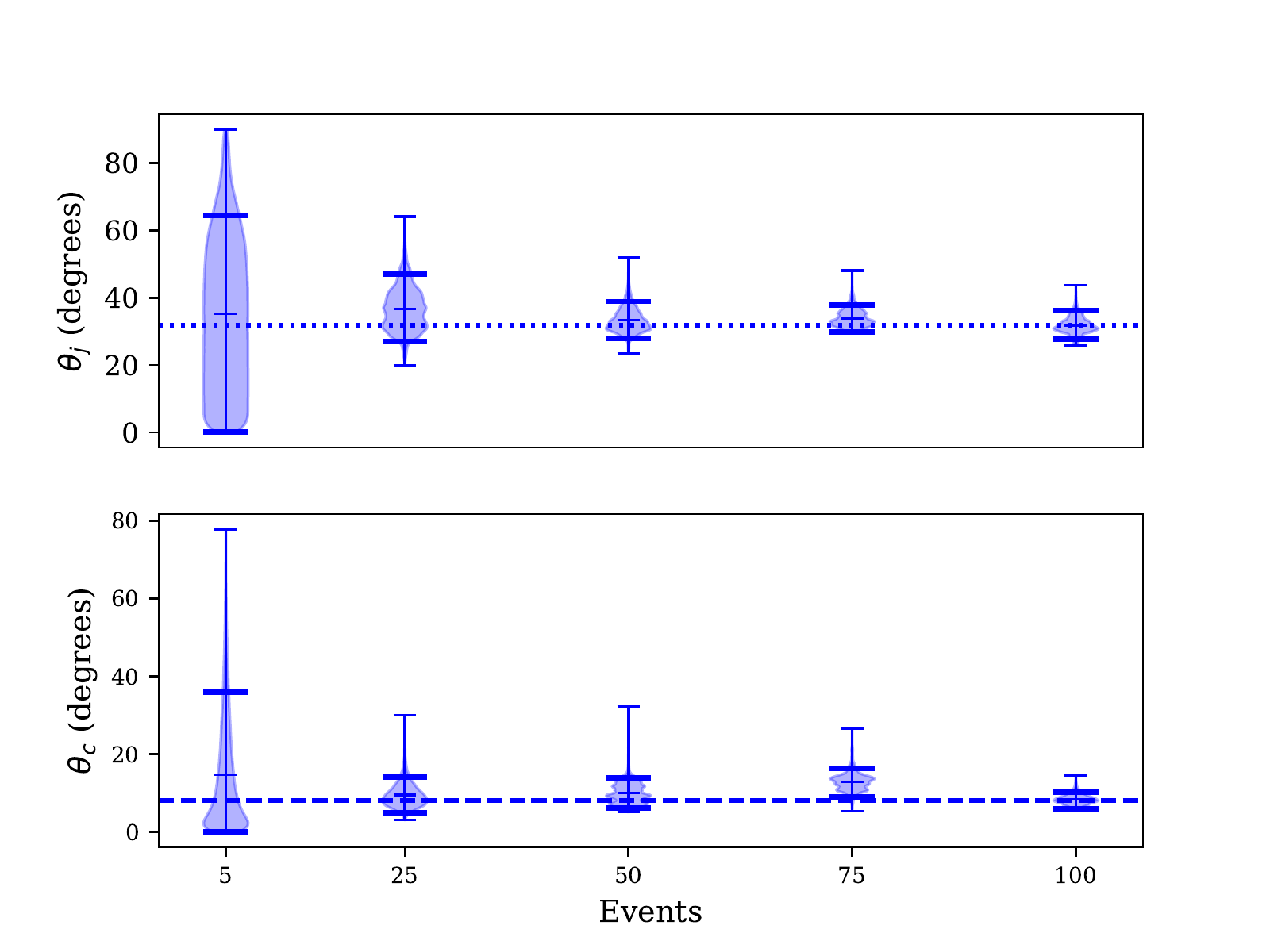}{0.5\textwidth}{(b) $p(\theta_{\text{c}}|\mathcal{D}_{\text{PL}}^{\ast},M=\text{PL},I)$ and $p(\theta_{\text{j}}|\mathcal{D}_{\text{PL}}^{\ast},M=\text{PL},I)$ \label{fig:PL_violins}}}
	%\gridline{\fig{./figures/Model_compare_results_GB_funcs.pdf}{0.5\textwidth}{(c)\label{fig:GB_funcs}}
	%\fig{./figures/Model_compare_results_PL_funcs.pdf}{0.5\textwidth}{(d)\label{fig:PL_funcs}}}
	\caption{(a) and (b) show the posteriors on the model parameters for the Gaussian jet model (GJ) and power-law model (PL) evaluated on different subsets of $\mathcal{D}_{\text{GJ}}^{\ast}$ and $\mathcal{D}_{\text{PL}}^{\ast}$ respectively. 
	The violin plots are bounded by the minimum and maximum of the posteriors and have widths indicative of the density of samples.
	The median value is marked by a narrow horizontal bar in the middle of each plot.
	The inner thick markers indicate the narrowest $90\%$ credible interval. 
	Generally the constraints on the parameters tighten when larger numbers of events are considered. 
	%The predicted form of $y(\theta)$ is then reconstructed for the Gaussian jet model in (c) and the power-law jet model in (d) from these data sets. 
	\label{fig:violins}}
\end{figure}

The posteriors of the jet structure model parameters are also determined. 
Figure~\ref{fig:violins} (left) displays the posteriors on $\theta_{\sigma}$ given one realization of $\mathcal{D}_{\text{GJ}}$ (denoted $\mathcal{D}_{\text{GJ}}^{\ast}$) with different number of events. 
The posterior distributions are contained between a minimum and maximum horizontal bar. 
The middle line indicates the median and the shaded area in the bounds illustrates the density of posterior samples.
The thick inner markers bound the narrowest $95\%$ credible interval. 
The injected value is marked by the purple dashed line. 
With more events considered, the posteriors close tighter around the injected value. 
However the parameters are well constrained even after 25 events.
%Figure~\ref{fig:violins} (lower left) displays $y(\theta)$ reconstructed from the $\theta_\sigma$ samples given $\mathcal{D}_{\text{GJ}}^{\ast}$. 
Similarly, Figure~\ref{fig:violins} (right) shows the posteriors on $\theta_{\text{c}}$ and $\theta_{\text{j}}$ given $\mathcal{D}_{\text{PL}}^{\ast}$.
% and the reconstructed $y(\theta)$ (lower right). 
Again, the posteriors tend to close around the injected values with an increased number of events. 
While the posteriors on $\theta_{\sigma}$ look Gaussian-like, the posteriors on $\theta_{\text{c}}$ and $\theta_{\text{j}}$ have a long tail towards high angles, however the modal values of the posteriors correspond well to the injected values. 
The values from Figure~\ref{fig:violins} are recorded in Table~\ref{tab:CIs}.

\section{Conclusion and Discussion}\label{sec:conclusion}

Modeling the jet structure of sGRBs will allow us to further understand events like GW170817/GRB$\,170817$A.
We demonstrate a comprehensive and fully Bayesian way of incorporating the distance and viewing angle posteriors from GW inference with the prompt emission from any accompanying on-axis or off-axis sGRB to examine the apparent jet structure of sGRBs.
Model comparison is performed between jet structure models and within 100 such events the power-law and Gaussian jet structure models can be distinguished with a significant log Bayes factor $> 5$ as shown in Figure~\ref{fig:model_comp_4}.
Despite the large widths of the GW posteriors, jet structure model parameters can be well constrained after just 25 events for both the Gaussian jet structure and the power-law shown in Figure~\ref{fig:violins} to within $90\%$ credible intervals of widths $ \Delta\theta_{\sigma}\sim2.8^{\circ} $, $ \Delta\theta_{\text{c}}\sim9.1^{\circ} $ and $ \Delta\theta_{\text{j}}\sim19.9^{\circ} $.
After 100 events the constraints become tighter still with bounds of $ \Delta\theta_{\sigma}\sim1.6^{\circ} $, $ \Delta\theta_{\text{c}}\sim4.2^{\circ} $ and $ \Delta\theta_{\text{j}}\sim8.5^{\circ} $ as seen in Figure~\ref{fig:violins} and Table~\ref{tab:CIs}.

The current BNS rate is predicted to be $110-3840\,\text{Gpc}^{-3}\,\text{yr}^{-1}$ \citep{abbott2019gwtc}.
With a search volume of $2.5\times10^{6}\,\text{Mpc}^{3}\,\text{yr}$, 
the number of BNS detections during the third observing run is predicted to be $ 2^{+8}_{-2}\,\text{yr}^{-1} $ and increase to $8^{+42}_{-7}\,\text{yr}^{-1}$ in the fourth observing run after KAGRA \citep{somiya2012detector} is operational and with further detector design improvements, increasing the search volume to $1.3\times10^{7}\,\text{Mpc}^{3}\,\text{yr}$ \citep{2019LRR....21....3A}. 
The analysis does not require every detection to be accompanied by a sGRB, however the sky localization of the BNS from the GW detection must be within the field of view of an operational GRB detector.
A fraction of these events will be shielded by the Earth or occur while satellites transition over the South Atlantic Anomaly and therefore would not qualify for this analysis.
Therefore a dataset of $ \sim 25 $ BNS events could be obtainable within this period but it may take until the third generation detectors are built until a dataset of $\sim 100$ BNS events is amassed.
With the installment of third generation GW detectors (eg. Einstein Telescope \citep{punturo2010einstein}), the BNS detection rate will likely increase into the thousands per year.
This will be complemented by developments in GRB detectors (eg. THESEUS \citep{amati2018theseus}) which will provide improved sensitivity in GRB observations and allow for a more detailed study of the jet structure when viewed outside the jet confines.

This work serves as a proof-of-principle and a starting point for a more inclusive analysis which can be applied to real data.
This would include a sGRB likelihood that accounts for the spectral models used in the fluence calculation such as described in \citet{gruber2014fermi,yu2016fermi}.
Currently the GW distance and inclination posteriors are approximated as normal distributions and any selection effects that are present in real data are neglected.
When real non-Gaussian GW posteriors are considered these effects will need to be accounted for.
A similar analysis is performed by \citet{farah2019counting}, where detection statistics are considered to account for such effects.

There are various ways in which the model described here could be expanded and assumptions could be removed.
The variation in observed isotropic equivalent energies of GRBs may not be to do with the jet structuring but instead due to large variations of the intrinsic energy of each sGRB's central engine. 
A wider prior on the face-on isotropic equivalent energy would account for this possibility, as \citet{fan2017probing} demonstrated for estimating the luminosity of sGRBs with uniform apparent jet structures. However this would require a much larger dataset to discern jet structure models of sGRBs with non-uniform apparent jet structure.
The current analysis could also be modified to account for variations in sGRB jet structure model parameters by placing priors on them and instead inferring the hyperparameters, such as performed by \citet{biscoveanu2019constraining}, who perform a similar analysis.
This would also allow for possible correlations between parameters to be distinguished such as the possible anti-correlation between $E_{\text{iso}}$ and $\theta_{\text{c}}$ \citep{nakar2019electromagnetic}.
We assume that the sGRB apparent jet structure can be modeled by a power-law or Gaussian jet structure.
However the Lorentz factor and gamma-ray production efficiency are likely much more dynamic than as assumed in this proof-of-principle analysis and may cause the apparent structure to differ from these simple models especially at large angles.
In future work we aim to perform model comparison on the intrinsic jet structure and to also model the varying Gamma factor of the source over viewing angle would give deeper insight into the underlying sGRB astrophysics.
This can be done by fully accounting for beaming effects by considering a model such as that described in \citet{ioka2019spectral}.
Whether or not every BNS merger produces a sGRB is still unknown.
\citet{lamb2016low} suggest that the majority of BNS mergers fail to produce a sGRB if their Lorentz factors are as low as other high-energy astrophysical phenomena such as blazars and active galactic nuclei.
The sGRB production rate from BNS mergers can be parameterized and incorporated similarly to \citet{williams2018constraints}.
While increasing the model complexity can help answer a number of interesting questions, it increases the required amount of data necessary to be able to perform model comparison.
Such a dataset will not be available until third generation GW detectors are operational.
Therefore until more data is available it is only feasible to make comparisons between specific models.

We focus on a power-law and Gaussian jet structure for the sGRBs, this analysis could expand to allow for comparison between any pairs of models. 
Often emission from the surrounding lower energy and Lorentz factor cocoon \citep{ramirez2002events,zhang2004propagation} has been modeled as a separate component from the central jet in a \textit{two-component} jet structure \citep{peng2005two,lazzati2019jet}.
There has also been much work into inferring the jet structure from the rise and decay of the observed afterglow (eg. \citet{lamb2017electromagnetic,lyman2018optical,troja2018outflow,wu2018constraining}), which could be used in tandem with this study to provide further evidence when comparing the different jet structure models.
This can include off-axis sGRB observations \citep{lamb2018grb,beniamini2020afterglow}, specifically through features of off-axis afterglows such as X-ray plateaus \citep{oganesyan2019structured,beniamini2019x}.

\acknowledgments

The authors would like to thank the referee for their insight.
The authors would also like to thank Gavin Lamb, Kentaro Mogushi and Francesco Pannarale for their helpful comments.
This work made use of the ARCCA Raven cluster, funded by STFC
grant ST/I006285/1 supporting UK Involvement in the Operation of Advanced LIGO.
F.H. was supprted by STFC grant number ST/N504075/1.
I.S.H. and D.W. was supported by STFC grant number ST/N005422/1.
J.V. was supported by STFC grant number ST/N005422/1 and partially supported by STFC grant number ST/K005014/2.


%% the manuscript Authors should list each code and include either a

\software{LALinference~\citep{veitch2015parameter}, matplotlib~\citep{Hunter:2007}, NumPy~\citep{oliphant2006guide}, PyMC3~\citep{salvatier2016probabilistic}, SciPy~\citep{2019arXiv190710121V}}

\begin{deluxetable*}{CCCCCCC}[h!]
	\tablecaption{Median $\pm 90\% $ credible intervals (degrees) of model parameters inferred from datasets $ \mathcal{D}_{\text{GJ}}^{\ast} $ and $\mathcal{D}_{\text{PL}}^{\ast}$ in Figure~\ref{fig:violins}. \label{tab:CIs}}
	\tablecolumns{7}
	\tablenum{1}
	\tablewidth{0pt}
	\tablehead{
		\colhead{Dataset} &
		\colhead{$\theta_{M}$} &
		\colhead{} & \colhead{} & \colhead{Events} & \colhead{} & \colhead{} \\
		\colhead{} & \colhead{} &
		\colhead{5} & \colhead{25} & \colhead{50} & \colhead{75} & \colhead{100} 
	}
	\startdata
	\mathcal{D}_{\text{GJ}}^{\ast} & \theta_{\sigma} & 11.9_{-8.7}^{+5.0} & 20.1_{-1.4}^{+1.4} & 19.4_{-1.4}^{+1.3} & 19.5_{-1.0}^{+1.0} & 19.4_{-0.8}^{+0.8} \\
	\hline
	\mathcal{D}_{\text{PL}}^{\ast} & \theta_{\text{c}} & 10.0_{-10.0}^{+25.9} & 9.0_{-4.1}^{+5.0} & 9.7_{-3.5}^{+4.2} & 13.0_{-3.9}^{+3.3} & 8.3_{-2.2}^{+2.0} \\
	& \theta_{\text{j}} & 34.2_{-34.2}^{+30.1} & 36.2_{-9.1}^{+10.8} & 32.7_{-4.9}^{+6.2} & 33.4_{-3.5}^{+4.4} & 31.4_{-3.8}^{+4.7} \\	\enddata
\end{deluxetable*}



\appendix

\section{Model Comparison}\label{sec:TI}

Given a dataset $\boldsymbol{\mathcal{D}}$ and two models $M_{1}$ and $M_{2}$, we can ask which model best describes the dataset by calculating the Bayes factor $\mathcal{B}_{1,2}$ defined as the ratio of the marginal likelihood of $M_{1}$ over $M_{2}$:

\begin{eqnarray}\label{equ:lnBF}
\mathcal{B}_{1,2} = \frac{p(\boldsymbol{\mathcal{D}}|M_{1},I)}{p(\boldsymbol{\mathcal{D}}|M_{2},I)},
\end{eqnarray}

where:

\begin{eqnarray}
p(\boldsymbol{\mathcal{D}}|M,I) =
\int
p(\boldsymbol{\mathcal{D}}|\boldsymbol{\Theta},\boldsymbol{\Phi},M,I)
p(\boldsymbol{\Theta},\boldsymbol{\Phi}|M,I)
d\boldsymbol{\Theta}d\boldsymbol{\Phi}.
\end{eqnarray}

This integral can be calculated using thermodynamic integration, a technique inspired by elements from statistical mechanics \citep{gelman1998simulating,lartillot2006computing}.
Let $p(\mathcal{D}|M,I)$ equate to the partition function of a model $M$, $Z=p(\mathcal{D}|M,I)$, and we denote $p(\mathcal{D}|\boldsymbol{\Theta},\boldsymbol{\Phi},M,I)
p(\boldsymbol{\Theta},\boldsymbol{\Phi}|M,I)$ to be state $q$. 
Therefore the probability of the system being in state $q$ is:

\begin{eqnarray}\label{equ:partition}
p=\frac{q}{Z},
\end{eqnarray}

which is equivalent to Bayes theorem with $p=p(\boldsymbol{\Theta},\boldsymbol{\Phi}|\mathcal{D},M,I)$, the joint posterior on $\boldsymbol{\Theta}$ and $\boldsymbol{\Phi}$. 
Now define:

\begin{eqnarray}\label{equ:temperature}
q_{\beta} = p(\mathcal{D}|\boldsymbol{\Theta},\boldsymbol{\Phi},M,I)^{\beta}
p(\boldsymbol{\Theta},\boldsymbol{\Phi}|M,I),
\end{eqnarray}

where $\beta$ is analogous to the thermodynamic definition $\beta\propto1/T$ where $T$ is the temperature. 
From this definition, $q_{0}=p(\boldsymbol{\Theta},\boldsymbol{\Phi}|M,I)$ and $q_{1}=p(\mathcal{D}|\boldsymbol{\Theta},\boldsymbol{\Phi},M,I)
p(\boldsymbol{\theta}|M,I)$. 
From Equation~\ref{equ:partition}, $Z_{0}=1$ and $Z_{1}=p(\mathcal{D}|M,I)$. 
Increasing $\beta$ then represents the transition from the `uninformed' prior state to the `informed' posterior state. 
The partition function holds the same properties as in thermodynamics, and is related to $q_{\beta}$ by the potential energy $U$:

\begin{eqnarray}
\frac{\partial \ln Z_{\beta}}{\partial\beta}=E_{\beta}[U]=E_{\beta}\left[\frac{\partial\ln q_{\beta}}{\partial\beta}\right].
\end{eqnarray}

An expression for $\ln p(\mathcal{D}|M,I)$ can be made:

\begin{eqnarray}
\ln p(\mathcal{D}|M,I) 
& = & 
\ln Z_{1} - \ln Z_{0}
= 
\int_{0}^{1}
\frac{\partial \ln Z_{\beta}}{\partial\beta}
d\beta \\
& = &
\int_{0}^{1}
E_{\beta}\left[\frac{\partial\ln q_{\beta}}{\partial\beta}\right]
d\beta \\
& = &
\int_{0}^{1}
E_{\beta}\left[
\ln p(\mathcal{D}|\boldsymbol{\Theta},\boldsymbol{\Phi},M,I)
\right]
d\beta,\label{equ:modelcompare}
\end{eqnarray}

from the definition of $q_{\beta}$. 
This integral can be approximated by calculating $E_{\beta}\left[
\ln p(\mathcal{D}|\boldsymbol{\Theta},\boldsymbol{\Phi},M,I)\right]$ at a range of discrete $0 \le \beta \le 1 $, where $\beta$ modifies the likelihood according to Equation~\ref{equ:temperature}.
We consider 25 temperatures, where $\beta$ is spread by $(i/24)^{-5}$ where $i=0,...,24$ according to \citet{calderhead2009estimating}.

%% The reference list follows the main body and any appendices.}{p(\mathbf{r}_{j},\boldsymbol{\theta}_{j}|\boldsymbol{\mathcal{D}},I)}



\bibliography{references}% Produces the bibliography via BibTeX.



\end{document}